\documentclass[english,showpacs,superscriptaddress]{revtex4}
\usepackage[T1]{fontenc}
\usepackage[dvips]{graphicx}

\makeatletter

\usepackage{babel}
\makeatother
\begin{document}

\title{Incompressible Turbulence as Nonlocal Field Theory}

\author{Mahendra K. Verma }

\email{mkv@iitk.ac.in}

\homepage{http://home.iitk.ac.in/~mkv}

\affiliation{Department of Physics, Indian Institute of Technology, Kanpur 208016,
INDIA}

\begin{abstract}
It is well known that incompressible turbulence is nonlocal in real
space because sound speed is infinite in incompressible fluids. The
equation in Fourier space indicates that it is nonlocal in Fourier
space as well. Contrast this with Burgers equation which is local
in real space. Note that the sound speed in Burgers equation is zero.
In our presentation we will contrast these two equations using nonlocal
field theory. Energy spectrum and renormalized parameters will be
discussed.
\end{abstract}

\pacs{47.27.Ak, 47.27Gs}

\maketitle

\section{Introduction}

Generic \textbf{}Equations in Physics, like Diffusion Equation, Schrodinger
Equation are local in real space. Take Schrodinger's equation for
example:\[
-i\hbar\frac{\partial\psi}{\partial t}=-\frac{\hbar^{2}}{2m}\nabla\psi+V(\mathbf{x})\psi,\]
where $V(\mathbf{x})$ is the potential, and $\psi(\mathbf{x},t)$
is the wavefunction. Clearly, to compute $\psi(\mathbf{x},t+dt)$
given $\psi(\mathbf{x},t)$, we need the local function, and finite
number of its derivatives. In this talk we investigate whether the
equations for fluid flows is local or not.

Fluid flows are described by Navier-Stokes (NS) equation, continuity
equation, and the equation of state given below:\begin{eqnarray}
\frac{\partial\mathbf{u}}{\partial t}+\left(\mathbf{u}\cdot\nabla\right)\mathbf{u} & = & -\frac{1}{\rho}\nabla p+\nu\nabla^{2}\mathbf{u},\label{eq:NS1}\\
\frac{\partial\rho}{\partial t}+\nabla\cdot\left(\rho\mathbf{u}\right) & = & 0,\label{eq:rho}\\
p & = & p(\rho),\label{eq:p}\end{eqnarray}
where $\mathbf{u},p,\rho$ are the velocity, pressure, and density
field respectively. $\nu$is the kinematic viscosity of the fluid.
We nondimensionalize the above equations by scaling the quantities
appropriately, e. g., divide $\mathbf{u}$ by velocity scale $U$.
Navier-Stokes equation gets modified to

\begin{eqnarray}
\frac{\partial\mathbf{u}}{\partial t}+\left(\mathbf{u}\cdot\nabla\right)\mathbf{u} & = & -\frac{C_{s}^{2}}{U^{2}}\nabla\rho+\frac{\nu}{UL}\nabla^{2}\mathbf{u}\label{eq:NS2}\end{eqnarray}
where $L$ is the length scale, and $C_{s}=\sqrt{{\frac{dp_{0}}{d\rho_{0}}}}$
is the Sound Speed. If the sound speed $C_{s}$ and $\frac{\nu}{UL}$
are finite, then we can find $\mathbf{u}(\mathbf{x},t+dt)$ and $\rho(\mathbf{x},t+dt)$
given $\mathbf{u}(\mathbf{x})$ and $\rho(\mathbf{x},t)$ ( assuming
that $\mathbf{u}(\mathbf{x},t)$,$\rho(\mathbf{x},t)$, and their
first and second derivatives are finite). For typical flows $\frac{\nu}{UL}$
is finite, so for finite $C_{s}$, Navier-Stokes equation is local
in real space. 

Note that the disturbances propagate with the sound speed. The larger
the sound speed, larger the range of influence per unit time. Still
the influence moves locally as long as the sound speed is finite.
When the sound speed is infinite, then disturbances can propagate
instantaneously, and all parts of the system starts interacting; hence
the system becomes nonlocal. Hence, the equations for fluid flows
become nonlocal when the sound speed is infinite, which is the case
for incompressible fluids $(\delta\rho=0)$. The speed of propagation
is infinite in Newton's law of gravitation as well as in Coulomb's
interactions between the charged particles. These are some other examples
of nonlocal interactions in physics. 

We can abstract the above physics using a 2D mesh of spring-mass system.
For finite spring constant, the disturbance propagates with a finite
speed, and the physics is local. When the spring constant is very
large, the physics is still local, but the range of propagation per
unit time becomes quite large. Here mass is pulled-pushed by local
spring only (4 of them). When the spring constant become infinite,
then whole system behaves like a solid and the speed of propagation
becomes infinite. This system has nonlocal interactions; the movement
of the mass at a point is affected by masses and springs at large
distances. In fact, in this nonlocal system, we can think of a mass
connected by all other masses, like in Coulomb's interactions or in
Calegaro-Sutherland model. 

Any real fluid has finite sound speed however large it may be. In
practice, the fluid is considered incompressible if $C_{s}/U\gg1$.
The properties of this fluid is expected to match with the ideal incompressible
fluid. This is based on an assumption that the properties of fluid
with $C_{s}\rightarrow\infty$ matches with $C_{s}=\infty$, or $(C_{s}\rightarrow\infty)=(C_{s}=\infty)$.
This assumption seems reasonable, but we are not aware of any strict
mathematical result showing this. Note that $(\nu\rightarrow0)\ne(\nu=0)$. 

In the next section we discuss incompressible Navier-Stokes equation.

\section{Incompressible Navier-Stokes}

Before we proceed further, we remark that for incompressible fluids
the normalized $-\nabla p=-\frac{C_{s}^{2}}{U^{2}}\nabla\rho$ is
finite even though $C_{s}$ is infinite. Also note that the normalized
term $\frac{\nu}{UL}\nabla^{2}\mathbf{u}$ is finite.

The continuity equation yields a constraint equation\[
\nabla\cdot\mathbf{u}=0,\]
the substitution of which in NS equation gives Poission's equation
for $p$,\[
\nabla^{2}p=-\nabla\cdot\left\{ \mathbf{u}\cdot\nabla\mathbf{u}\right\} \]
Therefore,\[
p(\mathbf{x},t)=\int\frac{-\nabla\cdot\left\{ \mathbf{u}\cdot\nabla\mathbf{u}\right\} }{\left|\mathbf{x}-\mathbf{x}'\right|},\]
which is the Coulomb's operator (nonlocal). Clearly the computation
of $p(\mathbf{x},t)$ and consequently that of $\mathbf{u}(\mathbf{x},t+dt)$
requires knowledge of $\mathbf{u}(\mathbf{x}')$ at all positions.
This is another way to infer that incompressible NS is nonlocal in
real space. Landau \cite{LandFlui:book}, Frisch \cite{Fris:book},
and others reached the above conclusion.

It is customary to study NS in Fourier space. Let us investigate whether
NS is local or nonlocal in Fourier space. NS equation in Fourier space
is 

\[
\frac{\partial u_{i}(\mathbf{k},t)}{\partial t}=-ik_{j}\int u_{j}(\mathbf{q})u_{i}(\mathbf{p})+ik_{i}\frac{k_{j}k_{m}}{k^{2}}\int u_{j}(\mathbf{q})u_{m}(\mathbf{p})\]
with $\mathbf{k=p+q}$. Note that the second term arises due to pressure. 

Since $u_{i}(\mathbf{k},t)$ requires knowledge of $u_{j}(\mathbf{q})u_{i}(\mathbf{p})$
where $\mathbf{p}$ and \textbf{$\mathbf{q}$} could be very different
from $\mathbf{k}$, hence NS is nonlocal interactions in Fourier Space.
If we interpret NS in terms of energy transfer, we find that the energy
is exchanged between the all the three modes of the triad. Kraichnan
\cite{Krai:59} and Dar et al. \cite{Dar:flux} devised formulas to
measure the strength of interactions in fluid turbulence. In this
paper we will use Dar et al.'s \emph{mode-to-mode formalism} \cite{Dar:flux}
in which the energy transfer rate from Fourier mode $\mathbf{p}$
to Fourier mode $\mathbf{k}$ with Fourier mode $\mathbf{q}$ acting
as a mediator is given by\begin{equation}
S(\mathbf{k|p|q)}=Im[\mathbf{k}\cdot\mathbf{u}(\mathbf{q})\mathbf{u}(\mathbf{p})\cdot\mathbf{u}(\mathbf{k})].\label{eq:S}\end{equation}
The above quantity can be computed using numerical simulation or using
analytic tools. Earlier, Domaradzki and Rogallo \cite{Doma:Local2}
and Waleffe \cite{Wale} calculated the above using EDQNM approximation.
Recently Verma et al. \cite{Ayye} calculated the above using field-theoretic
technique. In this paper we will report analytical result obtained
using first-order field-theoretic calculation. In this scheme, under
the assumption of homogeneity and isotropy we obtain\begin{equation}
\left\langle S(k'|p|q))\right\rangle =\frac{T_{1}(k,p,q)C(p)C(q)+T_{2}(k,p,q)C(k)C(q)+T_{3}(k,p,q)C(k)C(p)}{\nu(k)k^{2}+\nu(p)p^{2}+\nu(q)q^{2}},\label{eq:Savg}\end{equation}
where $T_{i}$'s are functions of $k,p$ and $q$. To save space,
we have skipped all the details for which the reader is referred to
Verma et al. \cite{Ayye}.

\begin{figure}
\includegraphics[%
  scale=0.5]{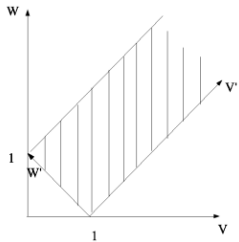}

\caption{\label{Fig:vw} The interacting triad $(\mathbf{k,p,q})/k=(1,v,w)$
under the condition $\mathbf{k}=\mathbf{p+q}$ is represented by a
point $(v,w)$ in the hatched region. The axis $(v',w')$ are inclined
to axis $(v,w)$ by 45 degrees. Note that the local wavenumbers are
$v\approx1,w\approx1$ or $v'\approx w'\approx1/\sqrt{{2}}$. }
\end{figure}

We focus our attention on the inertial range where the interactions
are self-similar. Therefore, it is sufficient to analyze $S(k'|p|q)$
for triangles $(1,p/k,q/k)=(1,v,w)$. Since, $|k-p|\le q\le k+p$,
$|1-v|\le w\le1+v$, hence any interacting triad $(1,v,w)$ is represented
by a point $(v,w)$ in the hatched region of Fig. \ref{Fig:vw} \cite{Lesl:book}.

For convenience, $\left\langle S(v',w')\right\rangle $ are represented
in terms of new variables $(v',w')$ measured from the rotated axis
shown in the figure \ref{Fig:vw}. It is easy to show that $v=1+(v'-w')/\sqrt{{2}},w=(v'+w')/\sqrt{{2}}$.

\begin{figure}
\includegraphics[%
  scale=0.5]{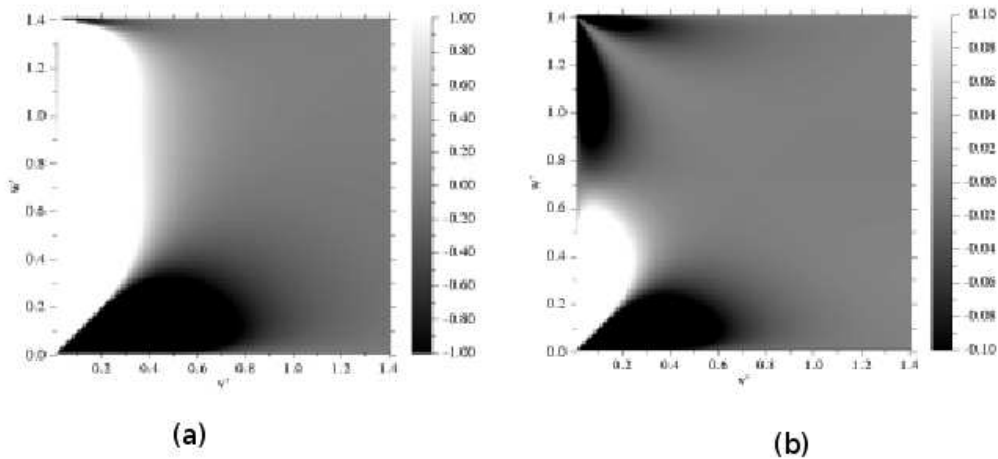}

\caption{\label{Fig:S} Density plot of $\left\langle S(v',w')\right\rangle $
of Eq. (\ref{eq:Savg}) without the bracketed factor for (a) 3D (b)
2D. }
\end{figure}

The local wavenumbers are $v\approx1,w\approx1$, while the rest are
called nonlocal wavenumbers. We substitute $C(k)$ and $\nu(k)$ in
Eq. (\ref{eq:Savg}) and rewrite $S(k|p|q)$ in terms of $v',w'$.
For details refer to Verma et al. \cite{Ayye}. Fig. \ref{Fig:S}
illustrates the density plots of $\left\langle S(v',w')\right\rangle $.
Fig. (a) shows the plot for 3D, while Fig. (b) shows the one for 2D.
We can draw the following conclusions from the plots.

\begin{enumerate}
\item When $v'\rightarrow1$ or $v\rightarrow0$, we find that $S(k|p|q)$
is large positive for 3D and large negative for 3D. This shows that
the nonlocal interactions are strong.
\item The value of $S$ at $(v,w)=(1,1)$, or $(v',w')=(1/\sqrt{{2}},1/\sqrt{{2}})$
is zero in both 2D and 3D. When $v\approx w\approx1$, $S$ is small
indicating that local interactions are small. 
\item When $v\rightarrow0$, $S>0$ for 3D and $S<0$ for 2D. This is reminiscent
of forward cascade in 3D, and backward cascade in 2D.
\end{enumerate}
Hence, we find that the interactions in the incompressible fluid turbulence
is nonlocal. This result appears to contradict Kolmogorov's phenomenology
which predicts local energy transfer in Fourier space. We will show
below that the shell-to-shell energy transfer rates are local even
though the interactions are nonlocal.

\section{Shell-to-shell energy transfers in turbulence.}

The wavenumber space is divided into shells $(k_{0}s^{n},k_{0}s^{n+1})$,
where $s>1$, and $n$ can take both positive and negative values.
The energy transfer rate from $m$th shell $(k_{0}s^{m},k_{0}s^{m+1})$
to $n$th shell $(k_{0}s^{n},k_{0}s^{n+1})$ is given by \cite{Dar:flux}\begin{equation}
T_{nm}=\sum_{k_{0}s^{n}\leq k\leq k_{0}s^{n+1}}\sum_{k_{0}s^{m}\leq p\leq k_{0}s^{m+1}}\left\langle S(k|p|q)\right\rangle .\label{eq:shell}\end{equation}
 If the shell-to-Shell energy transfer rate is maximum for the nearest
neighbours, and decreases monotonically with the increase of $|n-m|$,
then the shell-to-shell energy transfer is said to be local.

$T_{nm}$ can be computed using numerical simulations or using analytical
tools. Zhou \cite{Zhou:Local} calculated similar quantity. In the
following we plot $T_{nm}$ obtained using numerical simulation \cite{Ayye}.
Clearly, \emph{shell-to-shell energy transfer is local as envisaged
by Kolmogorov.}

\begin{center}%
\begin{figure}
\includegraphics[%
  scale=0.3]{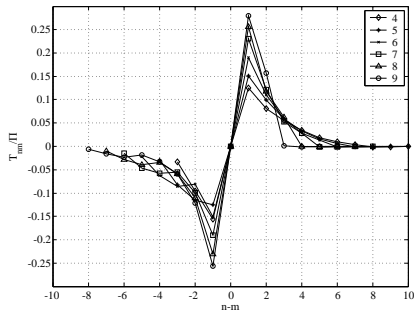}

\caption{\label{Fig:T-sim} Plots of normalized shell-to-shell energy transfer
$T_{nm}/\Pi$ vs. $n-m$ for $m=4..9$. The plots collapse on each
other indicating self-similarity.}
\end{figure}
\end{center}

We \cite{Ayye} have also computed the shell-to-shell energy transfer
rates using field-theoretic method. The reader is referred to the
original paper for the details. The plots of $T_{nm}$ for both 3D
and 2D fluid turbulence given below.

\begin{figure}
\includegraphics[%
  scale=0.5]{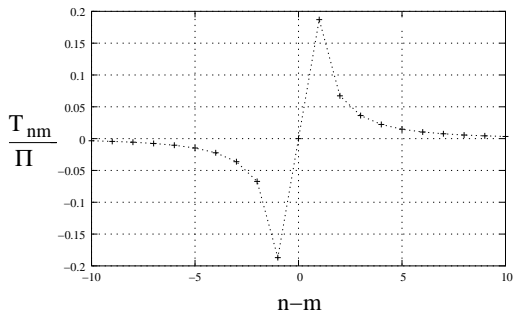}

\caption{\label{Fig:shell3D} Plot of normalized shell-to-shell energy transfer
$T_{nm}/\Pi$ vs $n-m$ for $d=3$. The $n$th shell is $(k_{0}s^{n},k_{0}s^{n+1})$
with $s=2^{1/4}$. The energy transfer is maximum for $n=m\pm1$,
hence the energy transfer is local. The energy transfer is also forward.}
\end{figure}
From the above plot we can clearly deduce that energy transfer in
3D fluid turbulence is local. In fact, the values obtained from analytical
calculations match very well with the numerical values shown in Fig.
\ref{Fig:T-sim}.

We have done similar analysis for 2D fluid turbulence. The result
is shown below:

\begin{figure}
\includegraphics[%
  scale=0.5]{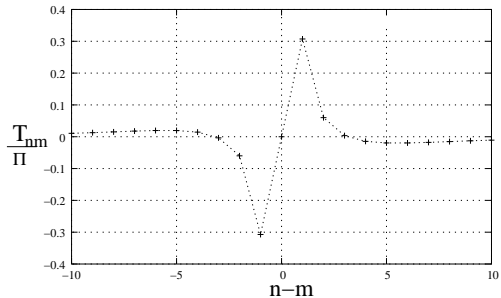}

\caption{\label{Fig:shell2D} Plot of normalized shell-to-shell energy transfer
$T_{nm}/\left|\Pi\right|$ vs $n-m$ for $d=2$ in the inertial range.
The energy transfer rate from the shell $m$ to the shells $m+1,$
$m+2$, $m+3$ is forward, but $m+4$ onward it is negative. The net
effect of all these transfer is the inverse energy flux $\Pi$. }
\end{figure}
The shell-to-shell energy transfer rates to the nearby shells are
forward, whereas the transfer rates to the far off shells are backward.
The net effect is a negative energy flux. This theoretical result
is consistent with Dar et al.'s numerical finding \cite{Dar:flux}.
The inverse cascade of energy is consistent with the backward nonlocal
energy transfer in mode-to-mode picture {[}$S(k|p|q)${]} (see Fig.
\ref{Fig:S}). Verma et al. \cite{Ayye} have shown that the transition
from backward energy transfer to forward transfer takes place at $d_{c}\approx2.25$. 

To summarize, the triad interactions in incompressible fluids is nonlocal
both in real and Fourier space. However, the shell-to-shell energy
transfer is local in Fourier space. Verma et al. \cite{Ayye} attribute
this behaviour to the fact that the nonlocal triads occupy much less
Fourier space volume than the local ones.

\section{Fully compressible limit: Burgers equation}

Let us go back to Eq. (\ref{eq:NS2}) and take the limit $C_{s}=0$.
This is the fully compressible limit, and the resulting equation was
first studied by Burgers. This equation, given below, is known as
Burgers equation (strictly speaking in 1D).\begin{eqnarray*}
\frac{\partial\mathbf{u}}{\partial t}+\left(\mathbf{u}\cdot\nabla\right)\mathbf{u} & = & \nu\nabla^{2}\mathbf{u}.\end{eqnarray*}
Clearly this equation is local in real space. What about in Fourier
space?

In Fourier space, the above equation is given by\[
\frac{\partial u_{i}(\mathbf{k},t)}{\partial t}=-i\int u_{j}(\mathbf{q})p_{j}u_{i}(\mathbf{k-q})-\nu k^{2}u_{i}(\mathbf{k}),\]
which implies that the interactions are nonlocal in Fourier space.
Note that the pressure term is absent in the above equation. 

The field-theoretic treatment of the above equation is rather complex
for arbitrary dimension. Here we attempt the self-consistent field-theoretic
treatment of one-dimensional Burgers equation for 1D Burgers equation.
In 1D, the energy spectrum of Burgers equation is given by \begin{equation}
C(k)=A\frac{\mu^{2}}{L}k^{-2},\label{eq:Burg:Ck}\end{equation}
where $L$ is the length of the system, $\mu$ is the shock strength,
and $A$ is a constant. Using dimensional arguments, we write the
renormalized viscosity of the following form \cite{McCo:book,MKV:MHD_RG}:

\begin{eqnarray}
\nu(k_{n}k') & = & \nu_{*}(k')\mu\sqrt{{\frac{A}{L}}}k_{n}^{-3/2}.\label{eq:Burg:nu}\end{eqnarray}
Unfortunately straight-forward application of self-consistent 
Renormalization Group (RG) procedure
of McComb \cite{McCo:book,MKV:MHD_RG} does not work. The contributions
of $<u^{>}(p)u^{>}(q)>$ is negligible; one needs to come up with
a cleverer renormalization scheme to obtain the renormalized viscosity.

To make a connection with Kolmogorov's theory of fluid turbulence,
we rewrite Eq. (\ref{eq:Burg:Ck}) as

\[
C(k)=A[\Pi(k)]^{2/3}k^{-5/3}\]
with the flux function $\Pi(k)$ as\begin{equation}
\Pi(k)=\frac{\mu^{3}}{L^{3/2}}k^{-1/2}.\label{eq:Burg:Pi}\end{equation}
Note that the flux $\Pi$ has become $k$-dependent. Verma \cite{MKV:KPZ}
and Frisch \cite{Fris:book} have shown that the flux function follows
a multifractal distribution. 

Question is whether we can compute the flux using field-theoretic
method. Since Burgers equation is compressible, the formula $S(K|p|q)$
is not applicable \cite{MKV:PR}. However we can still write the flux
using Kraichnan's combined energy transfer formula \cite{Krai:59}.
The energy flux crossing a wavenumber $k_{0}$ is given by\begin{eqnarray*}
\Pi(k_{0}) & = & \int_{k>k_{0}}dk\int_{p<k_{0}}\frac{1}{2}\Im\left[-ku(p)u(q)u(k)\right],\end{eqnarray*}
with $k+p+q=0$. We apply first-order perturbative method assuming
$u(k)$ to be quasi-normal as in fluid turbulence. We also make a
change of variable to \[
k=\frac{k_{0}}{u};\,\,\, p=\frac{k_{0}v}{u};\,\,\, q=\frac{k_{0}w}{u}.\]
To first order,\begin{eqnarray*}
\Pi(k_{0}) & = & \frac{k_{0}^{2}}{2}\int_{-1}^{1}du\int_{-u}^{u}(k)\frac{kC(p)C(q)+pC(k)C(q)+qC(k)C(p)}{\nu(p)p^{2}+\nu(q)q^{2}+\nu(k)k^{2}}\\
 & = & \frac{A^{3/2}}{\nu_{*}}\Pi(k_{0})\int_{-1}^{1}dv\,2(1-v^{1/2})\frac{w^{-2}+v^{-2}-(vw)^{-2}}{1+v^{1/2}+w^{1/2}},\end{eqnarray*}
with $w=1-v$. We find that the above integral converges and is equal
to 2.45. Hence,\[
1=\frac{A^{3/2}}{\nu_{*}}2.45.\]
Thus we show that $C(k),\nu(k),\Pi(k)$ given by Eqs. (\ref{eq:Burg:Ck},
\ref{eq:Burg:nu}, \ref{eq:Burg:Pi}) are consistent solution of 1D
Burgers equation. Note however that the renormalization group analysis
of Burgers equation is somewhat uncertain.

The spectral index of Burgers equation ($-2$) is very different from
the the spectral index of incompressible fluid turbulence ($-5/3$).
The difference arises due to the neglect of $-\nabla p$ term in Burgers
equation. The compressible effects are different in these two equations.
Burgers equation is local real space, while incompressible NS is nonlocal
in real space. 

It is interesting to compare the above results with Noncommutative
field theory, where the nonlocal interactions are included using parameter
$\theta$. Burgers equation is local, while incompressible NS is nonlocal
due to $-\nabla p$ term. Note that the $\nabla p$ term is nonlocal
in Coulomb's operator sense $(V\sim1/r)$. We are not aware of field-theoretic
ideas applied to Coulomb;s operator, which is one of the most important
operator in physics. We hope this investigation and its connection
with fluid turbulence will be carried out in future.

\begin{acknowledgments}
The above work is a result of collaborative work and discussions with
Arvind Ayyer, Shishir Kumar, Amar V. Chandra, V. Eswaran, and G. Dar.
\end{acknowledgments}

\end{document}